\RequirePackage{fix-cm}
\documentclass[smallcondensed]{svjour3} 

\usepackage{graphicx}
\usepackage[utf8]{inputenc}
\usepackage{float}
\usepackage[english]{babel}

\usepackage[a4paper,top=3cm,bottom=2cm,left=3cm,right=3cm,marginparwidth=1.75cm]{geometry}

\usepackage{amsmath}
\usepackage{graphicx}
\usepackage[colorinlistoftodos]{todonotes}
\usepackage[colorlinks=true, allcolors=blue]{hyperref}

\usepackage{sidecap}
\usepackage{caption}
\usepackage{amsmath}
\usepackage{braket}
\usepackage{multirow}
\usepackage{physics}

\title{Commissioning of the J-PET detector in view of the positron annihilation lifetime spectroscopy}
\author{
K. Dulski
\and
C.~Curceanu
\and
E.~Czerwi{\'n}ski
\and
A.~Gajos
\and
M.~Gorgol
\and
N.~Gupta-Sharma
\and
B.~C.~Hiesmayr
\and 
B.~Jasi{\'n}ska
\and
K.~Kacprzak
\and
Ł.~Kapłon
\and
D.~Kisielewska
\and
K.~Klimaszewski
\and
G.~Korcyl
\and
P.~Kowalski
\and   
N.~Krawczyk
\and
W.~Krzemie{\'n}
\and
T.~Kozik
\and
E.~Kubicz
\and
M.~Mohammed
\and    
Sz.~Nied{\'z}wiecki
\and
M. Pa{\l}ka
\and
M.~Pawlik-Nied{\'z}wiecka
\and
L.~Raczy{\'n}ski
\and
J.~Raj
\and
K.~Rakoczy
\and
Z.~Rudy
\and
S.~Sharma
\and
Shivani
\and
R.~Y.~Shopa
\and
M.~Silarski
\and
M.~Skurzok
\and
W.~Wi{\'s}licki
\and
B.~Zgardzi{\'n}ska
\and
P.~Moskal
}

\begin{document}
\institute{
K.~Dulski, kamil.dulski@gmail.com \at \and K.~Dulski \and E.~Czerwi{\'n}ski \and A.~Gajos \and N.~Gupta-Sharma \and K.~Kacprzak \and L.~Kaplon \and D.~Kisielewska \and G.~Korcyl \and T.~Kozik \and N.~Krawczyk \and E.~Kubicz \and M.~Mohammed 
\and Sz.~Nied{\'z}wiecki \and 
M.~Pa{\l}ka \and M.~Pawlik-Nied{\'z}wiecka \and J. Raj \and K.~Rakoczy \and Z.~Rudy \and S.~Sharma \and Shivani \and M.~Silarski  \and M.~Skurzok \and P.~Moskal \at Faculty of Physics, Astronomy and Applied Computer Science, Jagiellonian University,  S.~Łojasiewicza 11, 30-348 Kraków, Poland\label{WFAIS}
    \and 
C.~Curceanu \at
    INFN, Laboratori Nazionali di Frascati CP 13,  Via E. Fermi 40, 00044, Frascati, Italy\label{LNF}
    \and
    B.~C.~Hiesmayr \at
    Faculty of Physics, University of Vienna  Boltzmanngasse 5, 1090 Vienna, Austria\label{Vienna}
    \and
    M.~Gorgol \and B.~Jasi{\'n}ska \and B.~Zgardzi{\'n}ska \at
    Department of Nuclear Methods, Institute of Physics, Maria Curie-Sklodowska University, Pl.~M.~Curie-Sklodowskiej~1, 20-031 Lublin, Poland\label{UMCS}
    \and
    K.~Klimaszewski \and P.~Kowalski \and L.~Raczy{\'n}ski \and R.~Y.~Shopa \and W.~Wi{\'s}licki \at
    Department of Complex Systems, National Centre for Nuclear Research,  05-400 Otwock-Świerk, Poland\label{SWIERK}
    \and
    W.~Krzemie{\'n} \at
    High Energy Department, National Centre for Nuclear Research,  05-400 Otwock-Świerk, Poland\label{SWIERKHEP}
}

\maketitle
\date{Received: date / Accepted: date}
\maketitle

\begin{abstract}
\noindent 
The Jagiellonian Positron Emission Tomograph (J-PET) is the first PET device built from plastic scintillators. It is a multi-purpose detector designed for medical imaging and for studies of properties of positronium atoms in porous matter and in living organisms.
In this article we report on the commissioning of the J-PET detector in view of studies of positronium decays. 
We present results of analysis of the positron lifetime measured in the porous polymer.
The obtained results prove that J-PET is capable of performing simultaneous imaging of the density distribution of annihilation points as well as positron annihilation lifetime spectroscopy.\\
\end{abstract}
\\
\begin{small}
\textbf{Keywords} J-PET; Positronium; PALS; Positron Annihilation Lifetime Spectroscopy; TOT; XAD4.
\end{small}

\section{Introduction}
Positronium (Ps) is a purely leptonic atom constituting an excellent tool for studies of various phenomena~\cite{Casiddy1,Cassidy2} such as: discrete symmetries of nature in the leptonic sector~\cite{Ref5,Ref6,Ref7,Moskal4}, gravity effects on anti-matter~\cite{Gravi}, search for mirror photons~\cite{Mirror} or quantum entanglement~\cite{Beatrix,Nowakowski}. Recently it was also proposed that measurement of properties of positronium produced in the human body during the routine PET imaging,
may be used as diagnostic indicators complementary to the thus far applied standard uptake value~\cite{Morpho,PMBSubmit}. Such studies are carried out with the J-PET which was designed with a view to the imaging of the metabolism rate (via density distribution of annihilation points)~\cite{Moskal1,Moskal2,Moskal3} and with a view to studies of properties of positronium atoms in vacuum, in porous materials~\cite{Jasin} and in  living organisms~\cite{Morpho,PMBSubmit,Jasin1,Jasin2}.

\begin{figure}
\centering
\includegraphics[width=0.4\textwidth]{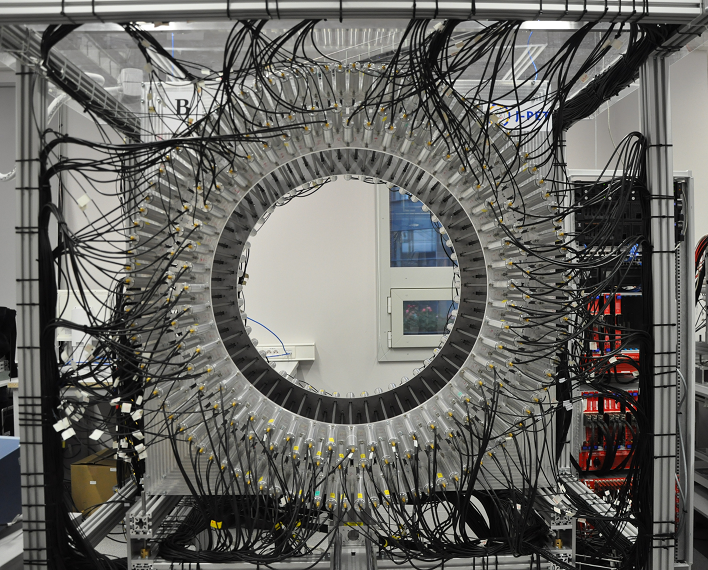}
\includegraphics[width=0.4\textwidth]{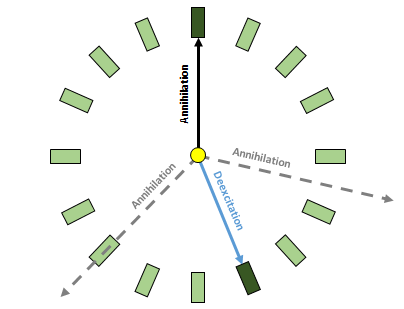}
\caption{\label{fig:JPET} (left) Photo of the J-PET detector.\\
(right) Cross section of one detection layer. For simplicity only 16 (out of 48) scintillator strips are shown. The superimposed arrows indicate photons from $\mbox{o-Ps} \to 3\gamma$ decay (black solid and dashed lines) and from deexcitation of $^{22}$Ne$^*$ (blue solid line). In the presented example only one annihilation photon (solid arrow) is registered.
}
\end{figure}

The J-PET detector (shown in Fig.~\ref{fig:JPET} (left)) is built from 192 plastic scintillators read out at two ends by photomultipliers~\cite{Moskal1,Moskal2,Moskal3,Szym}. The signals from photomultipliers are sampled in voltage domain with the newly designed, solely digital, front-end electronics~\cite{Palka}. Next, they are collected by the triggerless data acquisition system~\cite{Korcyl1,Korcyl2} and analyzed by means of the dedicated analysis framework~\cite{Krzemien}. The capabilities of the J-PET detector for the PET imaging were estimated using the Monte-Carlo simulations~\cite{Kowalski1,Kowalski2,Kowalski3} and proven experimentally by imaging of $^{22}$Na sources arranged in various configurations~\cite{Szym,Monika}. In this article we show the first lifetime spectrum of positronium atoms measured with the J-PET tomograph. The spectrum was determined for the positron annihilation in porous polymer XAD4. From the obtained lifetime distribution we determine fractional production intensities of para-positronium, direct anihilations, and ortho-positronium. The lifetimes and production intensities determined with the J-PET are compared to the results obtained for the XAD4 material by means of the standard positron annihilation lifetime spectroscopy (PAL) spectrometer~\cite{Jasin}. 
The lowest energy states of positronium, with orbital angular momentum equal to zero (L~=~0), possess significantly different mean-lifetimes depending on the spin. In vacuum the para-positronium (\textbf{S} = 0) mean lifetime (125~ps) and the mean lifetime of ortho-positronium (\textbf{S} = 1) (142~ns) differs by three orders of magnitude~\cite{Bader}. The latter, due to pick-off and ortho-para conversion processes~\cite{Pick_off,Convers,Zgardz} is shortened significantly when ortho-positronium is produced and trapped in the voids in the material. The lifetime distribution gives information about the size of free volumes in the material~\cite{Tao,Tao2}. Hence the PALS technique is widely used to characterize nanostructural pores in inorganic and organic materials e.g. such as porous silicas or polymers~\cite{PALS1,PALS2,PALS3,PALS4,PALS5}.
Recently a novel method was proposed for the in-vivo simultaneous determinations of the metabolic and positronium lifetime image~\cite{Morpho,PMBSubmit}. As a first step of the development of this method, the reconstruction of the density distributions of ortho-positronium annihilations into three photons was elaborated and validated with the Monte-Carlo simulations~\cite{Gajos,Daria}. In this article we present the first step en route to experimental validation of the PALS technique with the J-PET detector.

\section{Measurement and results}
In this article we report on the first measurement of positronium atoms performed with the J-PET detector in the period from October 14$^{th}$ to November 6$^{th}$ 2016.
The experiment was conducted with the $^{22}$Na sources wrapped in a $8~\mu$m thin Kapton foil surrounded with porous polymer XAD4\footnote{https://www.sigmaaldrich.com, CAS Number 9003-69-4}~\cite{Jasin}.
Such prepared source was placed inside a vacuum chamber at the centre. During the measurement a vacuum of about 1~Pa was sustained in the chamber.

In the standard PALS technique the positron lifetime spectra are measured typically with two detectors where one is set to register the prompt gamma photon with energy of about 1274~keV from the deexcitation of $^{22}$Ne$^*$ ($^{22}$Na $\to ^{22}$Ne$^* + e^+ + \nu \to ^{22}$Ne $+ \, \gamma + e^+ + \nu $) and the other detector is set to register the annihilation photons~\cite{Jasin,PALS1,PALS2} usually with energy of about 511~keV. Therefore, for the first tests of the J-PET in view of the PALS application we selected events with two signals for which one corresponds to the deexcitation photon and the other to the annihilation photon. An example of such event is presented in the right panel of Fig.~\ref{fig:JPET}. The J-PET is built from 192 separate detection modules, thus enabling us to measure the decay time (time between the annihilation and deexcitation) by 18336 different detector pairs.

\begin{figure}
\centering
\includegraphics[width=0.46\textwidth]{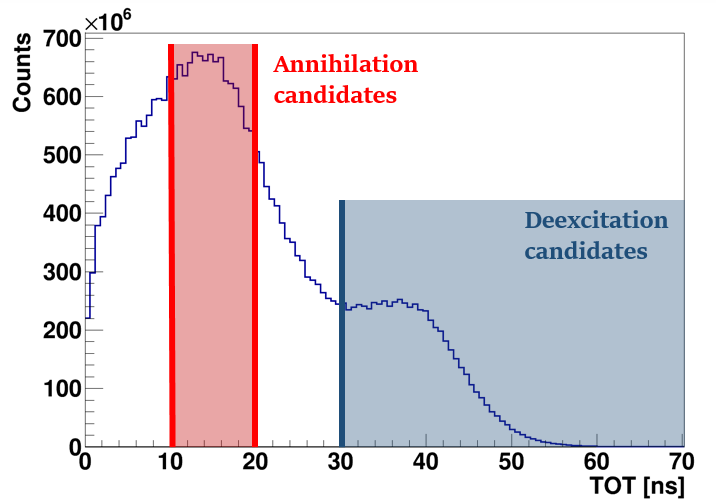}
\includegraphics[width=0.5\textwidth]{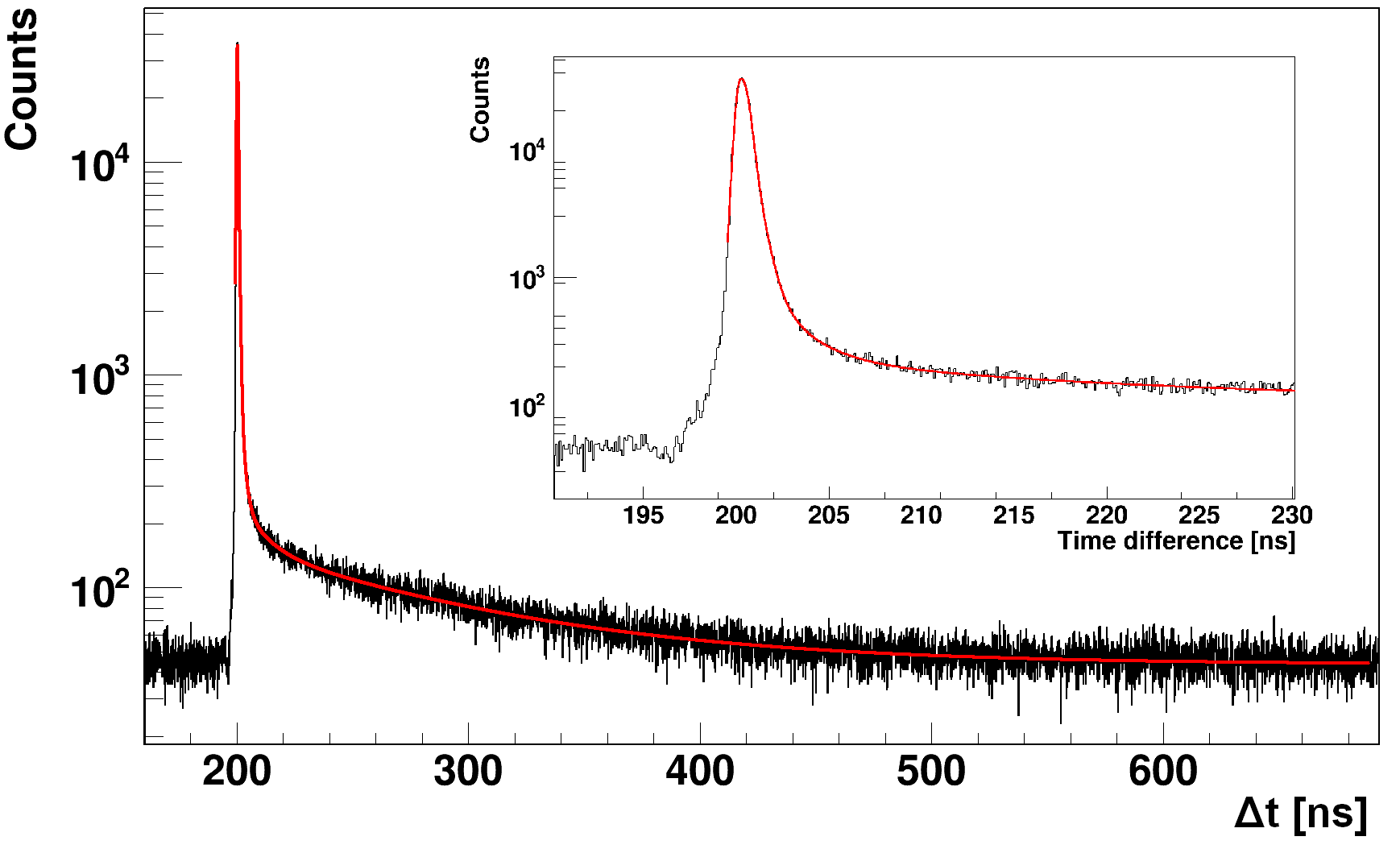}
\caption{\label{fig:Event} (left) Time over threshold distribution for the $^{22}$Na source surrounded with the XAD4 polymer as measured with the J-PET detector~\cite{Palka}. \\
(right) Positron lifetime distribution in the XAD4, obtained from measurement with the J-PET detector. Red line indicates result of the fit with six exponential components convoluted with two Gaussian distributions~\cite{Duls}.}
\end{figure}

In standard PALS crystal scintillators are used and 1274~keV and annihilation photons are identified by setting the energy loss windows corresponding to the total energy deposition (window around the photoelectric maximum). In the case of the J-PET detector, built from plastic scintillators, the interaction of annihilation and 1274~keV photons is predominantly due to the Compton effect and the photoelectric maximum is not visible. Moreover, in J-PET the charge of photomultiplier signals is not measured, and the energy loss is estimated based on the time over threshold (TOT) technique~\cite{Szym,Palka}.
A typical TOT distribution measured by J-PET is shown in the left panel of Fig.~\ref{fig:Event}. As expected, Fig.~\ref{fig:Event} reveals two overlapping Compton distributions with the Compton edges at about 22~ns and 45~ns corresponding to the annihilation and 1274~keV photons, respectively. One should emphasize that values of TOT are not changing linearly with the deposited energy~\cite{TOT}. The deexcitation photons may be identified when requiring that TOT value is larger than 30~ns.

The candidates for annihilation photons may be selected requiring that TOT value is less than 30~ns. However, for the purpose of this analysis, we select candidates for annihilation photon in the TOT range from 10~ns to 20~ns. Such choice minimizes the fake identification of annihilation photons due to the signals from the scattered deexcitation photon. A chosen restriction does not fully eliminate the background due to events in which both registered signals are due to primary and secondary interaction of the deexcitation photon.
In order to further decrease this kind of background, based on geometry we reject events, which with high probability originate from scatterings in the detector.
It is important to stress that the above described selection criteria are based on the reconstructed interaction position and TOT values and hence are independent of the measured time of photons interactions. Such method was chosen on purpose in order to avoid bias on the lifetime spectra due to event selection criteria.

After selection of events with one deexcitation and one annihilation photon (right panel of Fig~\ref{fig:JPET}) we determine the positron lifetime in the XAD4 for each event as a difference of times between interactions of annihilation and deexcitation photons ($\Delta t~=~t_a - t_d$) corrected for the time of flight of photon from the centre of the detector to the interaction point. Doing so, we assume that creation of the positron is equivalent to the deexcitation of the $^{22}$Ne$^*$. Such approach is justified taking into account, that the mean deexcitation time of $^{22}$Ne$^*$ is equal to about 3.7~ps~\cite{Firestone} and the thermalization time of positron in matter is in the order of tens of picoseconds~\cite{Kubica} only.  
The time of interactions was calculated taking into account calibration constants determined as described in Ref.~\cite{Magda}.
The positron lifetime distribution in XAD4, obtained by J-PET, is shown in the right panel of Fig.\ref{fig:Event}.
The spectrum shows a sharp maximum corresponding to the annihilation of para-positronium and direct $e^+e^-$ annihilations as well as long tail corresponding to the decays of ortho-positronium atoms. A superimposed line indicates the fit performed via the dedicated to J-PET fitting procedure described extensively in~\cite{Duls}. The fit function for the positron lifetime distribution is a convolution of the instrumental (resolution) function and the exponential decays, where linear coefficients are connected to probabilities of given components often referred as intensities. The exact formula is given in~\cite{Duls}. The lifetime and intensities of annihilations in the Kapton foil were fixed to 0.374 ns and 10\% respectively. The para-positronium mean lifetime was fixed to 0.125 ns as in~\cite{Jasin}.
The results of the fit are compared in Tab.~\ref{Tab:Results} with values quoted in reference~\cite{Jasin}. Results of the analysis carried out with both programs: J-PET program \cite{Duls} and LT~9.2  (commonly used for the PAL spectra analysis) \cite{LT} for the spectrum measured in J-PET are similar and allow to conclude that the J-PET measuring system is adapted to perform the positron lifetime measurements. The ortho-positronium lifetimes are in good agreement with the results expected on the basis of \cite{Jasin}. Discrepancies in the intensities of long-lived o-Ps component are due to the fact that relative acceptances of the J-PET detector for 2$\gamma$ and 3$\gamma$ events and energetic windows are different with respect to the standard PALS systems. In this article the selection was made based on the TOT values which exact relation is at present under evaluation. Therefore for the extraction of the proper absolute values of the intensities the relative measurement and selection sensitivities of the 2$\gamma$ and 3$\gamma$ annihilations have to be taken into account in the future quantitative analyses. 
\vspace{-0.7cm}
\begin{table}[h]
\centering
\caption{Results from the fitting positron lifetime distribution from the measurement with XAD4. Components with longest lifetimes are shown and compared to results obtained when applying the standard PALS technique~\cite{Jasin}. Uncertainties of parameters are putted in the brackets.}
\label{Tab:Results}
\begin{tabular}{|c|c|c|c|}
\hline
\textbf{Parameter} & Fit value by \cite{Duls} & Fit value by LT 9.2 &  Value from \cite{Jasin}\\
\hline
Lifetime of the 1$^{st}$ o-Ps component [ns] & 2.18 (10) & 2.12 (64) & 2.45 (25) \\
\hline
Intensity of the 1$^{st}$ o-Ps component [\%] & 3.5 (0.4) & 3.6 (0.9) & 3.3 (0.6) \\
\hline
Lifetime of the 2$^{nd}$ o-Ps component [ns] & 12.1 (0.7) & 11.3 (1.4) & 10.2 (0.6)\\
\hline
Intensity of the 2$^{nd}$ o-Ps component [\%] & 2.2 (0.2) & 2.5 (0.7) & 2.8 (0.5)\\
\hline
Lifetime of the 3$^{rd}$ o-Ps component [ns] & 94.1 (1.9) & 93.5 (1.5) & 90.8 (1.2)\\
\hline
Intensity of the 3$^{rd}$ o-Ps component [\%] & 21.9 (1.3) & 24.3 (2.2) & 44.8 (0.4)\\
\hline
Counts & \multicolumn{2}{|c|}{6.5 1E6} & 1.5 1E7 \\
\hline
FWHM of the first component [ns] & 0.628 (04) & 0.731 (09) & 0.474\\
\hline
Fraction of the first component [\%] & 93.5 (1.2) & 90.5 (1.0) & 16.1\\
\hline
FWHM of the second component [ns] & 0.351 (37) & 0.260 (22) & 0.295\\
\hline
Fraction of the second component [\%] & 6.5 (1.1) & 9.5 & 83.9\\
\hline
\end{tabular}
\vspace{-1.0cm}
\end{table}
\section{Summary}
We have presented the first positron life-time spectra measured with the J-PET detector. The obtained results prove capability of the J-PET for simultaneous PET imaging and PALS measurements.
\vspace{-0.5cm}
\begin{acknowledgements}
We are grateful to Steven Bass for correcting the final version of the manuscript.
This work was supported by The Polish National Center for Research
and Development through grant INNOTECH-K1/IN1/64/159174/NCBR/12, the
Foundation for Polish Science through the MPD and TEAM/2017-4/39 programmes, the National Science Centre of Poland through grants no.\
2016/21/B/ST2/01222,\linebreak[3] 2017/25/N/NZ1/00861,
the Ministry for Science and Higher Education through grants no. 6673/IA/SP/2016,
7150/E-338/SPUB/2017/1, 7150/E-338/M/2017 and 7150/E-338/M/2018, and 
the Austrian Science Fund FWF-P26783.
\end{acknowledgements}
\vspace{-0.5cm}
\bibliographystyle{plain}

\end{document}